# Longitudinal QSM: Enhancing consistency of multiple time point susceptibility maps via simultaneous reconstruction

Jiye Kim[1], Hwihun Jeong[2], Taechang Kim[1], Eunseon Jeong[3], Jinhee Jang[4], Yangsean Choi[3] and Jongho Lee[1,*]

[1]Laboratory for Imaging Science and Technology, Department of Electrical and Computer Engineering, Seoul National University, Seoul, Republic of Korea

[2]Department of Psychiatry, Brigham and Women's Hospital, Harvard Medical School, Boston, MA, USA

[3]Department of Radiology and Research Institute of Radiology, Asan Medical Center, University of Ulsan College of Medicine, Seoul, Republic of Korea

[4]Department of Radiology, Seoul St Mary's Hospital, College of Medicine, The Catholic University of Korea, Seoul, Republic of Korea

**Corresponding Author:**

Jongho Lee, Ph.D

Department of Electrical and Computer Engineering, Seoul National University

Building 301, Room 1008, 1 Gwanak-ro, Gwanak-gu, Seoul, Korea

Tel: 82-2-880-7310

E-mail: jonghoyi@snu.ac.kr

## Funding info

This work was supported by the National Research Foundation of Korea (NRF) grant funded by the Korea government (MSIT) (RS-2024-00349509, RS-2024-00438392), and by the Korea Health Industry Development Institute (KHIDI) (RS-2024-00439677).
This work was supported by the Institute of New Media and Communications, the Electric Power Research Institute, and the Institute of Engineering Research at Seoul National University, as well as Samsung Electronics (IO201216-08215-01).

## Abstract

Quantitative susceptibility mapping (QSM) has been increasingly applied in longitudinal studies of neurodegenerative diseases and aging to assess temporal alterations in brain iron and myelin. The accuracy of such investigations depends on the repeatability and sensitivity of measurements. However, the ill-posed nature of the QSM processing steps makes the reconstruction vulnerable to background field changes, head orientation changes, noise, and imperfect registration, which compromise repeatability and sensitivity and hinder reliable detection of true changes. To address these limitations, we propose Longitudinal QSM, a simultaneous reconstruction framework that jointly estimates susceptibility maps across time points while enforcing spatial sparsity of temporal changes. The method was evaluated through simulations and *in-vivo* experiments and compared with conventional reconstruction methods. Longitudinal QSM consistently reduced inter-scan variability and accurately recovered simulated lesion changes. Application to stroke patient and multiple sclerosis patient data further demonstrated that the framework stabilizes non-lesion variability while preserving lesion-related temporal changes. This approach offers a promising tool for monitoring subtle temporal changes in brain iron and myelin in various neurodegenerative diseases as well as throughout aging and development.

## 1. Introduction

Iron and myelin are essential components of the brain, and their abnormal accumulation or loss has been linked to various neurodegenerative diseases such as Parkinson’s disease (PD), multiple sclerosis (MS), and Alzheimer′s disease [1], [2].

Assessing their *in-vivo* spatial and temporal variations can therefore provide valuable insight into disease mechanisms and progression.

Quantitative susceptibility mapping (QSM) enables quantification of tissue magnetic susceptibility and provides a unique contrast sensitive to both paramagnetic and diamagnetic sources such as iron and myelin [3], [4]. QSM is reconstructed from multi-echo gradient-echo (GRE) data through processing steps including phase unwrapping, echo combination, background field removal, and dipole inversion. Since its introduction, QSM has been widely used to detect susceptibility sources related to neurological disorders [5], [6], [7], [8]. It has also been applied in longitudinal studies to assess temporal changes in brain iron and myelin associated with aging [9] and progression of neurodegenerative diseases such as MS and PD [10], [11], [12], [13], [14]. These applications highlight the potential of QSM for monitoring alterations in susceptibility over time.

Accurate longitudinal analysis, however, requires high repeatability and sensitivity to ensure that observed differences reflect genuine physiological alterations rather than measurement variability. QSM repeatability can be compromised by several factors, such as background susceptibility variation, head orientation change, dipole inversion consistency, noise, and imperfect registration [15], [16], [17], [18]. In particular, background susceptibility removal and dipole inversion are ill-posed processes that make the reconstruction highly sensitive to such confounding factors and can substantially affect reconstruction outcomes. Conventional longitudinal approaches, however, reconstruct each susceptibility map independently without accounting for these factors, thereby hindering accurate detection of longitudinal differences.

This study aims to develop a reconstruction method for longitudinal QSM data that improves inter-scan repeatability while preserving sensitivity to local susceptibility alterations. To this end, we introduce a simultaneous reconstruction framework that jointly estimates susceptibility maps across time points while accounting for background susceptibility variation and head orientation difference under a spatial sparsity constraint on susceptibility changes. Simulation experiments were conducted to evaluate the framework under background susceptibility changes, head rotations, and lesion susceptibility alterations. *In-vivo* experiments were performed in healthy volunteers to assess inter-scan repeatability, and in stroke and MS patients to demonstrate clinical feasibility.

## 2. Methods

### 2.1. Longitudinal QSM

Consider GRE measurements acquired at $N$ time points, $t_1$ (scan 1), $t_2$ (scan 2), ⋯, and $t_N$ (scan $N$). For each acquisition at $t_i$, we derive a local field map $f_i$, a magnitude image $M_i$, and a susceptibility map $\chi_i$, provided that all time point images are spatially coregistered.

The proposed method, Longitudinal QSM, assumes that true temporal susceptibility differences ($\chi_k - \chi_j$) are spatially sparse, meaning that temporal susceptibility changes are small and/or confined to localized regions such as lesions, while the rest of the brain remains stable. This assumption is incorporated into the Longitudinal QSM reconstruction by introducing an optimization framework (Figure 1) that simultaneously reconstructs susceptibility maps across all time points. The framework is formulated as follows:

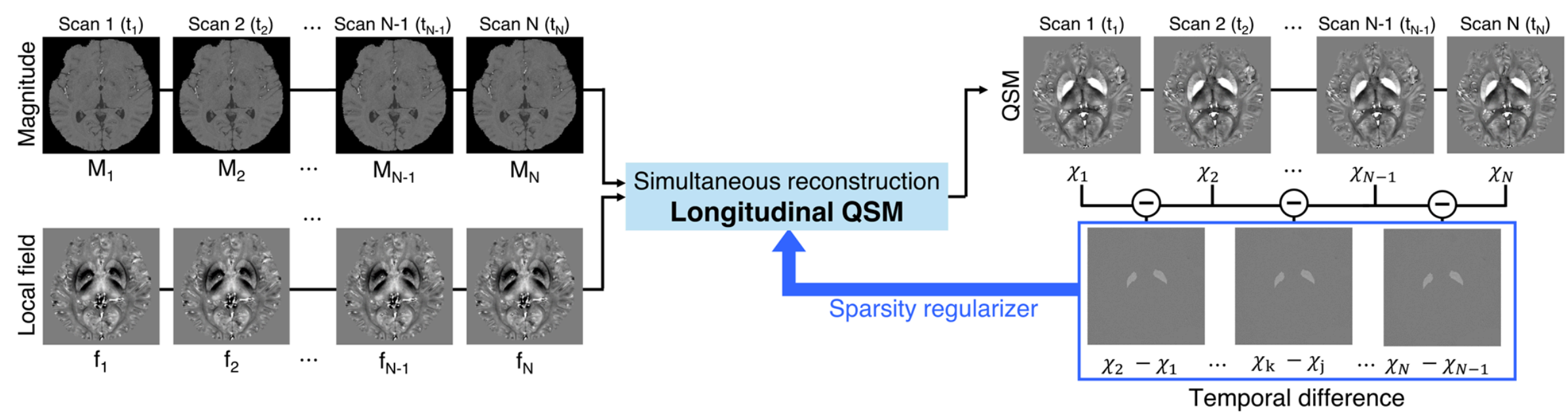

**Figure 1.** Overview of the proposed Longitudinal QSM framework for simultaneous reconstruction of longitudinal QSM data**.** Local field maps $\{f_i\}$ and magnitude images $\{M_i\}$ acquired at $i = 1, \cdots, N$ time points are jointly used to estimate the susceptibility maps $\{\chi_i\}$. The method assumes that temporal susceptibility differences $\{\chi_k - \chi_j\}$ are spatially sparse within the brain (i.e., small and/or confined to limited regions such as lesions). To incorporate this assumption, Longitudinal QSM minimizes a cost function consisting of (i) a sparsity regularizer applied to the temporal differences. While $\chi_1$ is constrained within the brain, $\chi_k - \chi_j$ can extend beyond the brain to capture background susceptibility variation, (ii) a dipole model-based data fidelity term, and (iii) a magnitude-driven edge-based structural regularizer.

$$\{\chi_1, \chi_2, \cdots, \chi_N\} = \arg \min_{\{\chi_1, \chi_2, \cdots, \chi_N\}} (\lambda_s \sum_{i=2}^{N} L^i_{sparsity} + \sum_{i=1}^{N} L^i_{dipole} + \lambda_{mag} \sum_{i=1}^{N} L^i_{mag}), \quad (1)$$

where $L^i_{sparsity}$, $L^i_{dipole}$, and $L^i_{mag}$ represent the temporal sparsity regularization term, dipole model fidelity term, and the magnitude-based structural regularization term for the *i*-th time point ($i = 1 \cdots, N$), respectively. Their explicit definitions are provided below. $\lambda_s$ and $\lambda_{mag}$ are regularization parameters controlling the weights of the sparsity and magnitude regularizers, respectively.

To incorporate the sparsity assumption of temporal susceptibility changes, we introduce a new sparsity prior on the susceptibility differences:

$$L^i_{sparsity} = \sum_{j=1}^{i-1} \|W_{brain}(\chi_i - \chi_j)\|_1, \; i = 2, \cdots, N. \quad (4)$$

In this equation, $W_{brain}$ denotes a common brain tissue mask excluding cerebrospinal fluid (CSF) [19] and major vessels [20], such that sparsity of susceptibility changes is primarily from brain tissue rather than CSF or large vessels. For each time point, a mask is derived from the corresponding magnitude image $M_i$, and the final $W_{brain}$ is defined as the intersection of masks across time points.

The dipole model fidelity term $L_{dipole}$ is designed to ensure consistency between the measured local field maps and those derived from the estimated susceptibility map:

$$L^i_{dipole} = \|W(f_i - D_i * \chi_i)\|_2^2 \quad (2)$$

Here, $D_i$ denotes the dipole kernel corresponding to the $B_0$ orientation of each scan, and $W$ is a brain mask.

The magnitude-based structural regularizer $L_{mag}$ is based on the Morphology Enabled Dipole Inversion (MEDI), one of the QSM dipole inversion approaches proposed by Liu et al. [21]. This term incorporates structural information from magnitude images to guide susceptibility reconstruction by promoting edge consistency between susceptibility and magnitude images:

$$L^i_{mag} = \|M_{i,G} \nabla \chi_i\|_1, \quad (3)$$

where $M_{i,G}$ is an edge-weighting mask derived from the gradient of the magnitude image, and $\nabla$ denotes the 3D gradient operator.

In the formulation, the susceptibility map at the first time point, $\chi_1$, is defined only within the brain mask $W$. The temporal differences $(\chi_k - \chi_j)$, however, are allowed both inside and outside the brain, while sparsity is enforced within the $W_{brain}$. This design accounts for background field components that may not be completely removed during background field removal. By permitting temporal differences outside the brain, residual background susceptibility variations can be jointly captured, preventing them from being misinterpreted as susceptibility changes within the brain.

The optimization problem is solved using an iterative reweighted least-squares (IRLS) scheme [22] combined with a block conjugate gradient solver. At each outer iteration, voxel-wise weights are updated from the current estimate, with $\epsilon = 10^{-6}$. With the weights fixed, the resulting equations are solved using a block conjugate-gradient method. The conjugate-gradient iterations are terminated when the relative residual falls below a threshold ($10^{-4}$), and the outer IRLS iteration is repeated five times. The susceptibility maps are initialized using susceptibility maps individually reconstructed with MEDI QSM [21].

## 2.2. Simulation experiment

To evaluate the performance of the proposed framework under controlled conditions, a simulation experiment was conducted using 1 mm isotropic COSMOS QSM [23] maps from four subjects in the $\chi$-sepnet dataset [24]. COSMOS QSM provides susceptibility maps by combining multi-orientation data and is commonly used as a reference standard in QSM studies [23]. We artificially introduced inter-scan variability by changing susceptibility distributions in the background and rotating head orientation, mimicking scans at two different time points. Two types of simulations were performed: (i) a scan–rescan simulation in which the brain susceptibility distribution was kept identical across scans, while background susceptibility variation and head rotation were introduced, and (ii) a lesion simulation in which localized susceptibility changes were additionally introduced within the scan-rescan simulation.

To change the susceptibility distribution in the background, two background susceptibility distributions were generated from the GRE data of one of the four subjects in the $\chi$-sepnet dataset using two distinct head orientations and were applied to all four subjects in the simulation. The GRE data acquired at the second orientation were registered to those of the first orientation prior to the processing [25]. For each orientation, the GRE phase data underwent phase unwrapping [26] and echo combination [27] to generate a total field map. A total susceptibility map was then reconstructed from the total field map using iLSQR QSM [28]. The background susceptibility distribution for each orientation was obtained by masking out the brain region in the total susceptibility map, thereby retaining only susceptibility sources outside the brain. Because brain masks differ across subjects, directly using the masked maps would introduce empty regions. Therefore, these empty regions were filled using nearest-neighbor interpolation to obtain spatially continuous background susceptibility maps. The two resulting distributions were denoted as $BKG_1$ and $BKG_2$. During simulation, $BKG_1$ and $BKG_2$ were re-masked using each subject's corresponding brain mask.

For the scan-rescan simulation, background susceptibility change and head rotation were applied to the COSMOS QSM reference: For scan 1, the local field map $f_1$ was generated by adding $BKG_1$ to the COSMOS map, followed by the dipole convolution and background field removal. For scan 2, $f_2$ was generated by adding $BKG_2$ with the rotated COSMOS map, followed by the same processing. Two rotation conditions were simulated: a nominal 3.4°, obtained by applying 2° rotations in all three axes (x, y, and z), and a nominal 6.8°, obtained by applying 4° rotations in all axes.

To demonstrate the sensitivity of the proposed framework to localized susceptibility alterations, lesion susceptibility changes were simulated by modifying only the scan 2 susceptibility maps. A globus pallidus ROI mask derived from the χ-separation atlas [29] was used to define the target region. Within this ROI, a constant susceptibility increment corresponding to 5% or 10% of the ROI mean susceptibility value was generated for each subject. Gaussian noise with a standard deviation of 0.003 ppm was added to the ROI to introduce realistic variability, followed by Gaussian smoothing (MATLAB *imgaussfilt*, $\sigma = 1$) to avoid sharp discontinuities. The resulting ROI susceptibility increment was added to the scan 2 susceptibility map, generating a lesion-modified susceptibility map, which served as the ground-truth susceptibility map for scan 2. Local field maps corresponding to these lesion-modified COSMOS maps were then generated using the same background susceptibility sources and rotation conditions described above.

For the proposed simultaneous reconstruction, the data were processed by Eq. 1 with regularization parameters $\lambda_{mag}$ = 0.0005 and $\lambda_s$ = 0.00025. For comparison, susceptibility maps were reconstructed individually using conventional MEDI QSM with $\lambda$ = 0.0005 and whole-brain reference.

The evaluation was performed by measuring voxel-wise repeatability using normalized root-mean-square error (NRMSE) and structural similarity index (SSIM) within the brain mask. Cerebrospinal fluid (CSF) [19] and blood vessels [20] were masked out when computing the metrics. The means and standard deviations of the metrics across the subjects were computed. In addition, ROI-wise susceptibility variation was calculated as

$$variation = \frac{\left| ROI\ mean\ (t_2) - ROI\ mean\ (t_1) \right|}{\left| ROI\ mean\ (t_1) \right|} \times 100\ (\%), \qquad (6)$$

across 22 anatomical regions defined in the $\chi$-separation atlas [29]. The repeatability metrics (NRMSE, SSIM, and ROI variation) were computed excluding the simulated lesion region. Within the lesion, sensitivity was assessed by comparing the assigned susceptibility changes with the measured changes using linear regression.

## 2.3. In-vivo experiment: Healthy controls

In-vivo experiments were performed on five healthy volunteers (mean age = 25.4 ± 1.0 years; four females and one male) to assess the inter-scan repeatability of the proposed method. Written informed consent was obtained from all participants in accordance with institutional review board (IRB) approval. All volunteers were scanned at 3T (Siemens Cima. X, Erlangen, Germany) using 3D multi-echo GRE with the following parameters: TR = 38.0 ms, TEs = 6.7 ms, 11.9 ms, 17.1 ms, 22.3 ms, 27.5 ms, and 32.9 ms, FOV = 256 × 224 × 160 $mm^3$, voxel volume = 1.0 × 1.0 × 1.0 $mm^3$, bandwidth = 287 Hz/pixel, and time of acquisition = 6.4 min. For each subject, three multi-echo GRE acquisitions were obtained: (i) a baseline scan, (ii) an immediate *repeat* scan without subject repositioning, and (iii) a *rescan* acquired after participant repositioning. Additionally, 3D magnetization-prepared rapid gradient echo sequence (TR = 2400 ms, TE = 2.1 ms, FOV = 256 × 256 × 224 $mm^3$ , voxel volume = 1.0 × 1.0 × 1.0 $mm^3$, bandwidth = 210 Hz/pixel, and time of acquisition = 5.45 min) was acquired for $T_1$-weighted image.

Processing of the multi-echo GRE data was performed as follows: A brain mask was first generated from the root-sum-of-squares magnitude image using the brain extraction tool in FSL [25]. The GRE phase data underwent phase unwrapping [26] followed by echo combination [27] and background field removal using V-SHARP [30], [31], resulting in a local field map for each scan. The second-echo magnitude images from the repeat and rescan sessions were registered to those of the baseline scan FLIRT in FSL [25]. The calculated transformation matrices were applied to their corresponding local field maps for registration. Longitudinal QSM reconstruction was then applied to both the scan–repeat and scan–rescan pairs to jointly reconstruct susceptibility maps across the two acquisitions with $\lambda_{mag}$= 0.0005 and $\lambda_s$ = 0.0015.

Scan-rescan repeatability was assessed using the same voxel-wise and ROI-wise quantitative metrics as in the simulation experiment (see Section 2.2), in comparison with that of MEDI QSM.

## 2.4. In-vivo experiment: Stroke patients

To demonstrate the clinical feasibility of the proposed framework, three stroke patients (mean age: 70 ± 12 years old; 2 females and 1 male) were scanned on a 3T MRI system (Siemens Vida, Erlangen, Germany). Two imaging sessions were performed per subject with an inter-scan interval of 13.4 ± 2.2 months. This retrospective study was approved by the IRB, and the requirement for informed consent was waived (KC26ZISI0017).

The 3D multi-echo GRE data were acquired with TR = 40.7 ms, TEs = 6.2 ms, 11.8 ms, 17.3 ms, 22.9 ms, FOV = 230 × 188 × 144 $mm^3$, voxel volume = 0.72 × 0.72 × 1.0 $mm^3$. Susceptibility-weighted images (SWI) were generated from multi-echo GRE data. Phase processing, background field removal, registration, and Longitudinal QSM reconstruction followed the same pipeline as described for the healthy volunteer dataset (see Section 2.3).

For evaluation, lesions identified on SWI were used for qualitative visual assessment of susceptibility maps between the two time points. The Longitudinal QSM reconstruction results were visually compared with susceptibility maps reconstructed individually using MEDI QSM.

## 2.5. In-vivo experiment: MS patients

Longitudinal data were acquired from 15 MS patients (8 females and 7 males; mean age at the first scan: 47.3 ± 12.7 years) using a 3T MRI (Siemens Vida, Erlangen, Germany). Among them, twelve had relapsing-remitting MS, and three had secondary progressive MS. Each patient underwent two imaging sessions separated by 13.4 ± 3.6 months. All patients signed a written consent form approved by the IRB.

The imaging protocol included 3D multi-echo GRE and 3D $T_2$-weighted FLAIR acquisitions. The 3D multi-echo GRE data were acquired with TR = 40.7 ms, TEs = 5.4 ms, 10.6 ms, 15.8 ms, 21.0 ms, 26.2 ms, and 31.4 ms, FOV = 208 × 156 × 120 $mm^3$, voxel volume = 0.5 × 0.5 × 2.0 $mm^3$, bandwidth = 250 Hz/pixel, and an acquisition time of 5.4 minutes. The 3D $T_2$-weighted FLAIR scans differed between the two sessions. First scan was acquired with TR = 5000 ms, TE = 328 ms, FOV = 230 × 230 × 160 $mm^3$, voxel volume = 1.0 × 1.0 × 1.0 $mm^3$, and bandwidth = 781 Hz/pixel. Second scan was acquired using TR = 4800 ms, TE = 343 ms, FOV = 256 × 256 × 176 $mm^3$, voxel size = 0.5 × 0.5 × 2.0 $mm^3$, and bandwidth = 651 Hz/pixel. Multi-echo GRE data processing, including spatial unwrapping, weighted echo combination, background field removal, and registration, was performed using the same procedures as in the healthy volunteer dataset (see Section 2.3). Longitudinal QSM reconstruction was also applied in the same way, using the same regularization parameters.

Lesions identified on T2-weighted FLAIR images were used to visually compare the susceptibility maps using the proposed framework and MEDI QSM.

## 2.5. Ablation study

In the proposed framework, susceptibility changes were jointly estimated across time points not only within the brain but also in the surrounding background region, while sparsity was enforced only within the brain. To examine the role of jointly modeling background susceptibility changes, an ablation study was conducted. In this ablation setting, temporal susceptibility differences ($\chi_k - \chi_j$) were restricted to the brain mask, with no temporal variation allowed outside the brain. All other components of the formulation, including the dipole fidelity term, magnitude-based regularization, and hyperparameters, were kept identical to those of the proposed method.

The ablation study was conducted in both the simulation experiment (Section 2.2) and the in-vivo healthy volunteer experiment (Section 2.3). Evaluation was performed using the same voxel-wise and ROI-wise metrics as described above, and the results were compared with those from the original Longitudinal QSM framework.

All data processing was performed using MATLAB (MATLAB 2024a, MathWorks Inc., Natick, MA, USA).

# 3. Results

## 3.1. Simulation experiment

Figure 2 shows the simulation results under background susceptibility variation and head rotation. When only background susceptibility variation is applied (Fig. 2A), both MEDI QSM and Longitudinal QSM exhibit minimal inter-scan differences. When head rotations are added to the background variation (3.4° rotation in Fig. 2B and 6.8° rotation in Fig. 2C), MEDI QSM exhibits increasingly pronounced artifacts, whereas Longitudinal QSM reduces these residuals and improves agreement between scan 1 and scan 2, as evident in the difference maps.

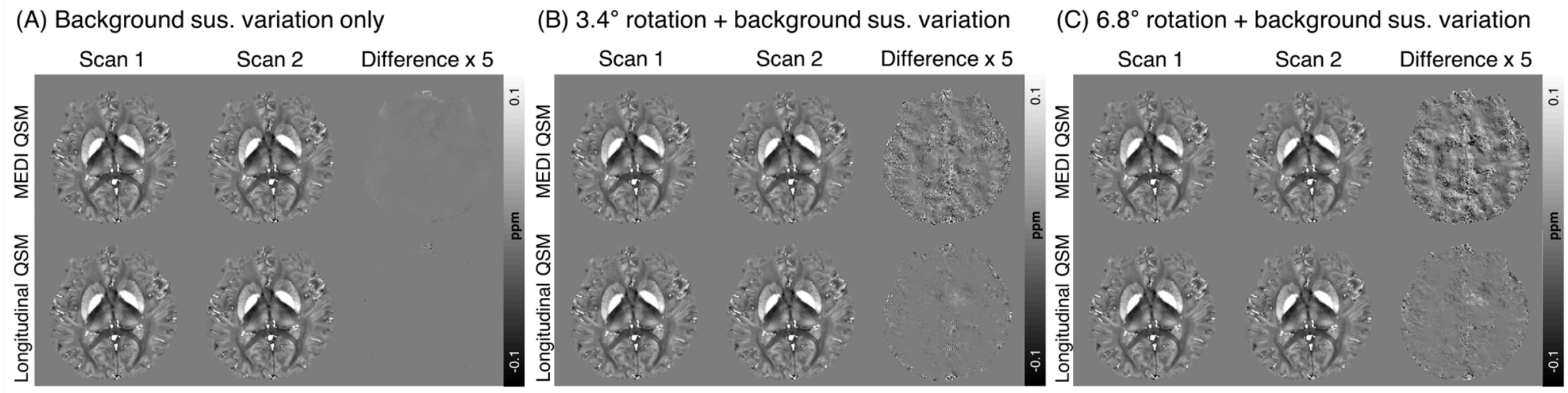

**Figure 2.** Simulation results demonstrating the effects of background susceptibility variation and head rotation. Three conditions are evaluated: (A) background susceptibility variation only (no rotation), (B) 3.4° rotation with background susceptibility variation, and (C) 6.8° rotation with background susceptibility variation. For each condition, susceptibility maps from MEDI QSM and Longitudinal QSM (Proposed) are shown for Scan 1 and Scan 2, along with the corresponding difference maps (×5 for visualization). MEDI QSM exhibits substantial inter-scan differences, which increase with rotation. Longitudinal QSM improves repeatability across all experimental conditions.

Voxel-wise repeatability metrics are summarized in Table 1. Compared with MEDI QSM, Longitudinal QSM consistently demonstrates lower NRMSEs and comparable or higher SSIMs across all conditions.

**Table 1.** Voxel-wise repeatability metrics in the simulation experiment for MEDI QSM and Longitudinal QSM across background susceptibility variation and rotation conditions. Lower NRMSE and higher SSIM indicate better repeatability.

| | Metric | Background susceptibility variation only | 3.4° rotation w/ background susceptibility variation | 6.8° rotation w/ background susceptibility variation |
|---|---|---|---|---|
| MEDI QSM | NRMSE (%) ↓ | 2.4 ± 0.3 | 23.1 ± 1.7 | 28.7 ± 2.2 |
| | SSIM ↑ | 0.999 ± 0.0001 | 0.988 ± 0.002 | 0.984 ± 0.003 |
| Longitudinal QSM | NRMSE (%) ↓ | 1.3 ± 0.1 | 11.5 ± 1.6 | 14.6 ± 1.9 |
| | SSIM ↑ | 0.999 ± 0.0001 | 0.998 ± 0.001 | 0.996 ± 0.001 |

Figure 3 summarizes ROI-wise susceptibility variation across 22 anatomical regions. Even with only small rotations and background susceptibility variation, MEDI QSM shows considerable fluctuations, with average variation approaching 4% and several regions exceeding this level. Such variability indicates that conventional processing steps alone can introduce substantial bias in longitudinal analyses. In contrast, Longitudinal QSM shows ROI variation of approximately 1% in most regions and conditions, with a maximum variation of 2.3%, highlighting its robustness to background susceptibility and head orientation changes.

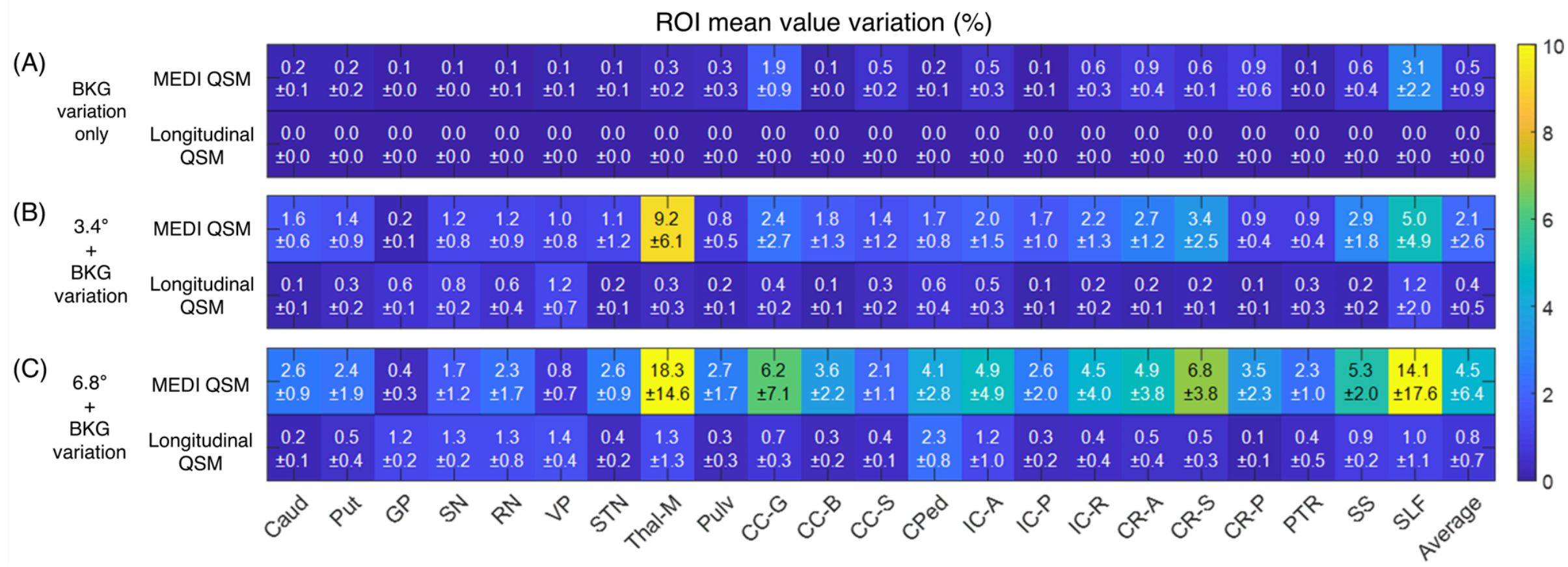

**Figure 3.** Simulation results illustrating ROI-wise susceptibility variation (%) under (A) background susceptibility (BKG) variation only, (B) 3.4° rotation with background susceptibility variation, and (C) 6.8° rotation with background susceptibility variation. Values represent mean ± standard deviation across four subjects for MEDI QSM and Longitudinal QSM. MEDI QSM shows substantial ROI-wise fluctuation, while Longitudinal QSM maintains variation around 1% or lower across nearly all regions, demonstrating improved inter-scan consistency. Caud: caudate nucleus, Put: putamen, GP: globus pallidus, NAc: nucleus accumbens, SN: substantia nigra, RN: red nucleus, VP: ventral pallidum, STN: subthalamic nucleus, Thal-M: medial thalamic nuclei, Thal-L: lateral thalamic nuclei, Pulv: pulvinar, CC-G: genu of corpus callosum, CC-B: body of corpus callosum, CC-S: splenium of corpus callosum, CPed: cerebral peduncle, IC-A: anterior limb of internal capsule, IC-P: posterior limb of internal capsule, IC-R: retrolenticular part of internal capsule, CR-A: anterior corona radiata, CR-S: superior corona radiata, CR-P: posterior corona radiata, PTR: posterior thalamic radiation, SS: sagittal stratum, SLF: superior longitudinal fasciculus.

The simulation results of the lesion susceptibility alteration in Figure 4 present a representative example with a 10 percent susceptibility increase in the globus pallidus. Both methods detect the simulated lesion. However, MEDI QSM maps exhibit pronounced variations both within the lesion region and across non-lesion areas. In contrast, Longitudinal QSM results preserve the lesion change while suppressing variations in non-lesion regions. Background susceptibility variations are also captured, as indicated by the red arrows.

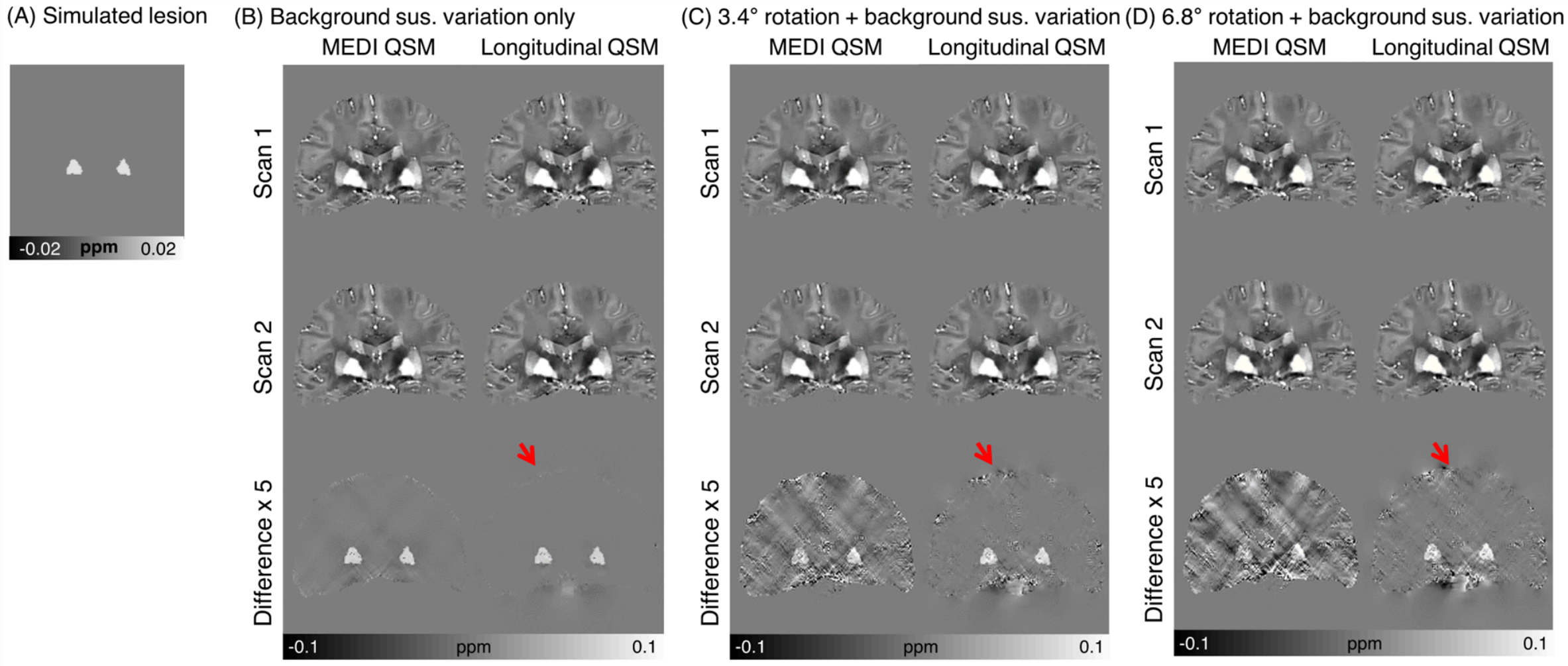

**Figure 4.** Simulated lesion experiment results demonstrating the sensitivity to localized susceptibility changes under background susceptibility variation and head rotation. (A) Simulated lesion generated by increasing the susceptibility of the globus pallidus by 10%. (B–D) Reconstructed susceptibility maps from MEDI QSM and proposed method for background susceptibility variation only (B), background susceptibility variation with 3.4° rotation (C), and background susceptibility variation with 6.8° rotation (D). Each row shows scan 1, scan 2, and the corresponding difference map (×5 for visualization). The proposed method accurately recovers the simulated lesion while suppressing non-lesion discrepancies and detects background susceptibility variation (red arrows). In contrast, MEDI QSM exhibits substantial residual differences with rotation, which obscure lesion depiction.

Quantitatively, under background susceptibility variation alone without rotation, NRMSE in non-lesion regions was 3.0 ± 0.4% for MEDI QSM and 1.3 ± 0.2% for Longitudinal QSM, while SSIM was similarly high for both methods (0.999 ± 0.0001). With 3.4° of rotation, inter-scan discrepancies increased in both methods but remained substantially lower for Longitudinal QSM, with NRMSE of 26.3 ± 1.8% for MEDI QSM and 11.9 ± 1.5% for Longitudinal QSM. Correspondingly, SSIM was 0.959 ± 0.007 for MEDI QSM and 0.999 ± 0.0002 for Longitudinal QSM. With 6.8° of rotation, the same trend was observed: NRMSE increased to 32.0 ± 2.2% for MEDI QSM and 15.1 ± 1.8% for Longitudinal QSM, while SSIM was 0.926 ± 0.01 for MEDI QSM and 0.999 ± 0.0005 for Longitudinal QSM. Overall, Longitudinal QSM demonstrates better repeatability across all conditions.

Consistently, Longitudinal QSM also reduced non-lesion ROI mean value variation. While MEDI QSM showed increasing average ROI variation with head rotation (0.8 ± 0.4%, 2.3 ± 1.0%, and 4.5 ± 3.4% for no rotation, 3.4°, and 6.8°, respectively), Longitudinal QSM maintained a lower average variation of 0.0 ± 0.0%, 0.4 ± 0.2%, and 0.8 ± 0.5% across the same conditions.

Figure 5 illustrates the agreement between the assigned and measured lesion susceptibility changes. Both methods exhibit underestimation, as indicated by slopes below 1. Longitudinal QSM shows less underestimation (slope = 0.915, $R^2$ = 0.96) compared with MEDI QSM (slope = 0.831, $R^2$ = 0.98). These findings confirm that Longitudinal QSM enhances both the detection of localized susceptibility changes and the stability of regional susceptibility estimates in the presence of background susceptibility change and rotation.

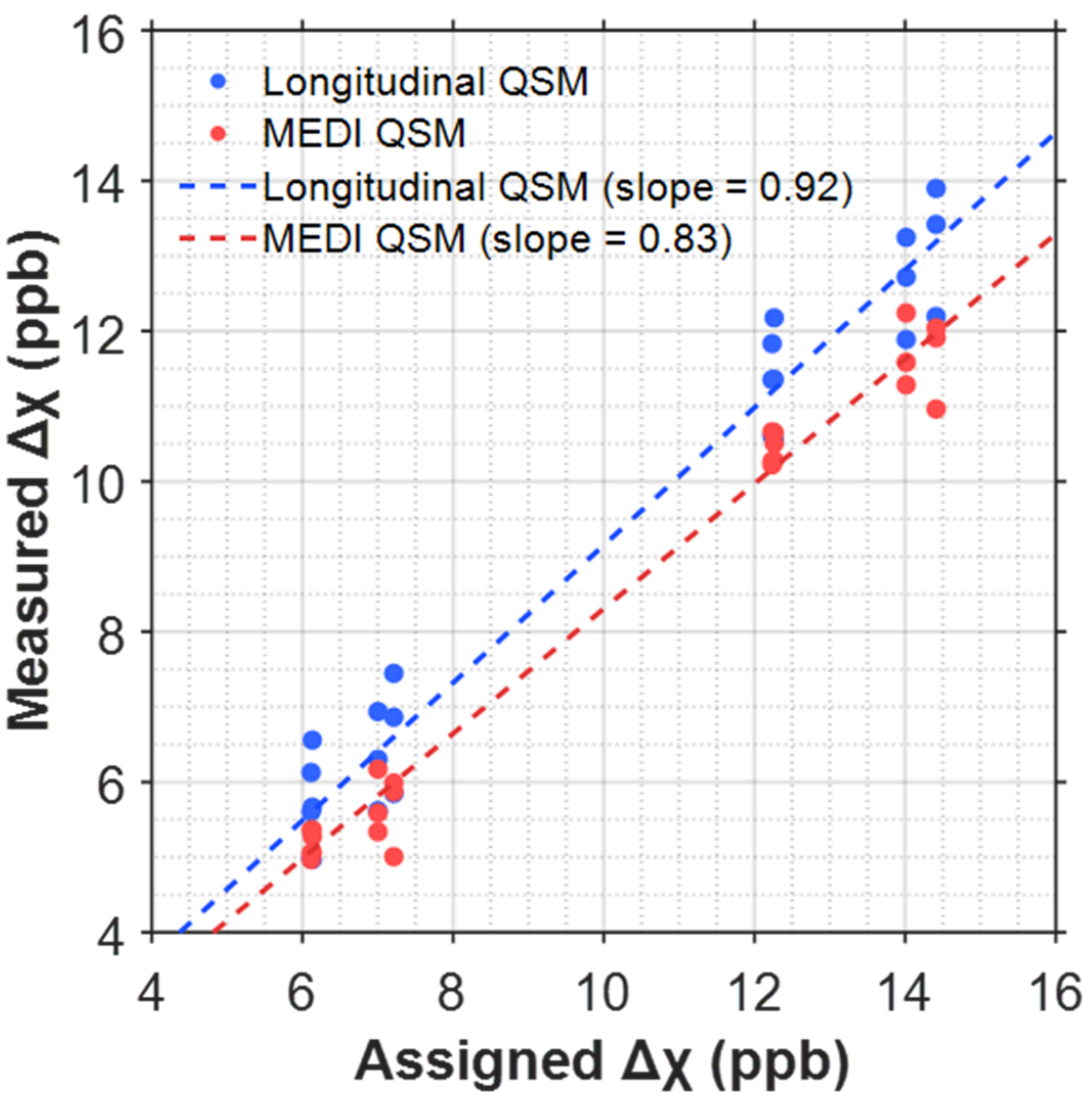

**Figure 5.** Simulated lesion experiment results of ROI analysis. Simulated lesion experiment results of linear regression analysis comparing the measured lesion susceptibility change ($\Delta\chi$) with the assigned ground-truth values across subjects and rotation conditions. Each point corresponds to one subject under one rotation setting. The proposed method (blue line) achieves a regression slope of 0.915 ($R^2$ = 0.96), indicating accurate estimation of susceptibility changes. MEDI QSM (red line) yields a lower slope of 0.831 ($R^2$ = 0.98), reflecting underestimation of lesion changes. These results demonstrate the improved sensitivity of the proposed method to localized susceptibility alterations.

## 3.2. In-vivo experiment: healthy volunteers

In the in-vivo healthy volunteer data, scan–rescan pairs exhibit larger changes in head orientation (4.8 ± 1.9°) than scan–repeat pairs (0.6 ± 0.3°). Correspondingly, inter-scan differences are more pronounced in the scan–rescan condition (Figures 6A and 6B). These inter-scan differences are reduced by the proposed method compared with MEDI QSM. Voxel-wise analysis also demonstrates that inter-scan differences are larger in the scan-rescan condition than in the scan-repeat condition. Nevertheless, Longitudinal QSM consistently reduced these differences compared with MEDI QSM. In the scan–repeat setting, NRMSE decreases from 46.0 ± 5.7% for MEDI QSM to 33.4 ± 3.8% for Longitudinal QSM, with SSIM increasing from 0.936 ± 0.01 to 0.975 ± 0.01. In the scan–rescan condition, although overall variability increases, NRMSE remains lower for Longitudinal QSM than for MEDI QSM at 41.4 ± 8.3% and 59.9 ± 12.1%, respectively, with a corresponding increase in SSIM from 0.893 ± 0.05 to 0.958 ± 0.03.

For the ROI-wise analysis, susceptibility variation is also reduced by Longitudinal QSM relative to MEDI QSM in both scan–repeat and scan–rescan conditions, as shown in Figures 6C and 6D.

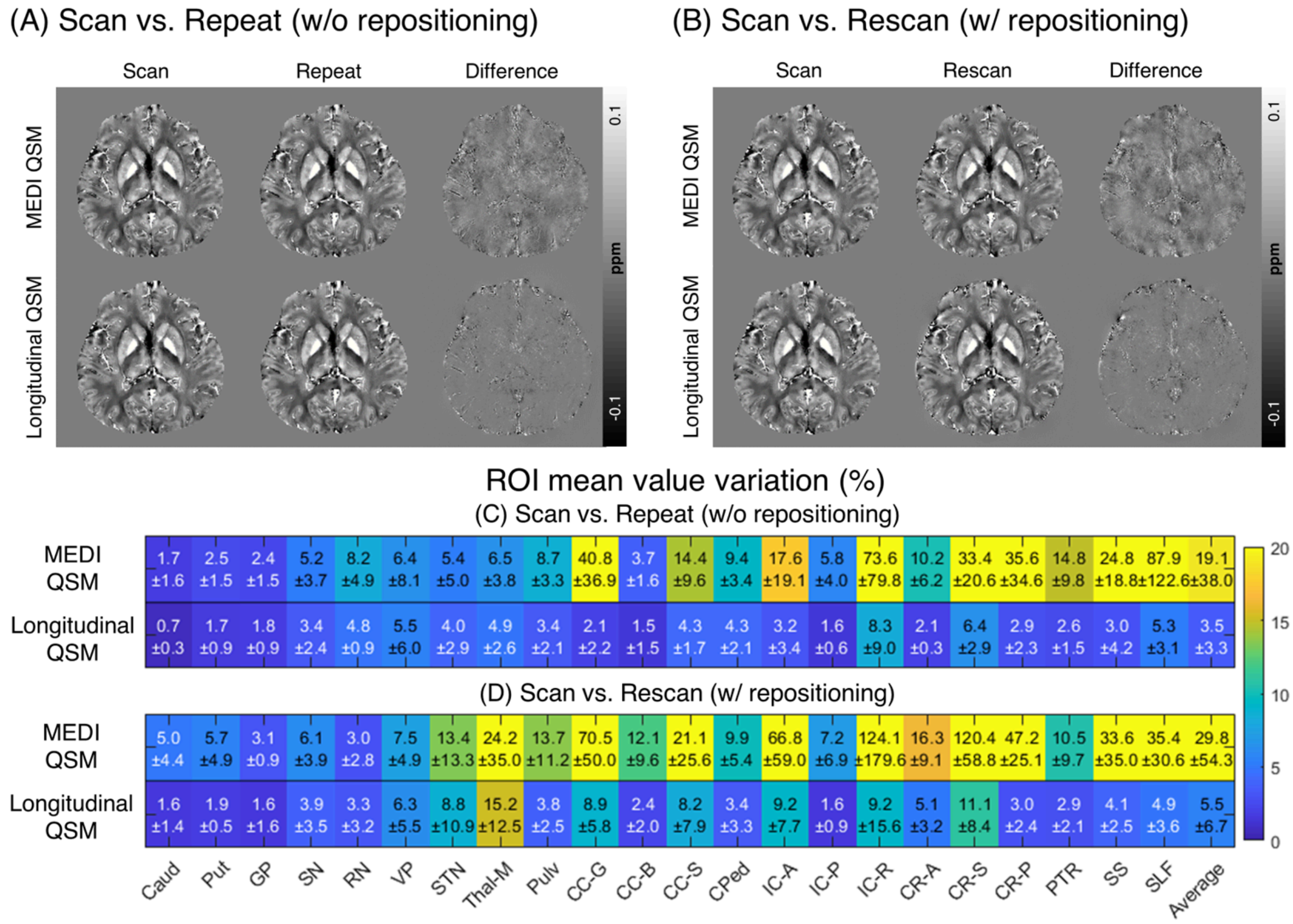

**Figure 6.** In-vivo results from healthy volunteers showing scan–repeat (A) and scan–rescan (B) comparisons. For each case, susceptibility maps reconstructed with MEDI QSM and the proposed

method are displayed along with the corresponding difference maps. In the scan–repeat pair (A), the proposed method reduces residual differences relative to MEDI QSM. In the scan–rescan pair (B), subject repositioning introduces more pronounced inconsistencies when processed with MEDI QSM, whereas Longitudinal QSM suppresses these non-biological differences and yields more consistent susceptibility maps. (C-D) ROI-wise susceptibility variation for in-vivo healthy volunteer experiments. (C) Scan vs. repeat and (D) scan vs. rescan comparisons are shown for 22 anatomical regions. MEDI QSM exhibits substantial inter-scan fluctuations, particularly in the scan–rescan condition, where several ROIs exceed 20%. Longitudinal QSM suppresses these ROI-wise differences across all regions in both scan–repeat and scan–rescan evaluations, providing better inter-scan reproducibility. Caud: caudate nucleus, Put: putamen, GP: globus pallidus, NAc: nucleus accumbens, SN: substantia nigra, RN: red nucleus, VP: ventral pallidum, STN: subthalamic nucleus, Thal-M: medial thalamic nuclei, Thal-L: lateral thalamic nuclei, Pulv: pulvinar, CC-G: genu of corpus callosum, CC-B: body of corpus callosum, CC-S: splenium of corpus callosum, CPed: cerebral peduncle, IC-A: anterior limb of internal capsule, IC-P: posterior limb of internal capsule, IC-R: retrolenticular part of internal capsule, CR-A: anterior corona radiata, CR-S: superior corona radiata, CR-P: posterior corona radiata, PTR: posterior thalamic radiation, SS: sagittal stratum, SLF: superior longitudinal fasciculus.

## 3.3. In-vivo experiment: Stroke patients

The results from a representative stroke patient are shown in Figure 7. On SWI, the lesion (orange arrow) is visible at both time points and appears slightly enlarged over the ten-month interval between scans. Although the lesion is identifiable on both scans, MEDI QSM shows widespread non-specific susceptibility variations that limit inter-scan consistency and hinder lesion depiction. In contrast, Longitudinal QSM suppresses non-lesion variations and provides a clearer depiction of the lesion, particularly in the second scan.

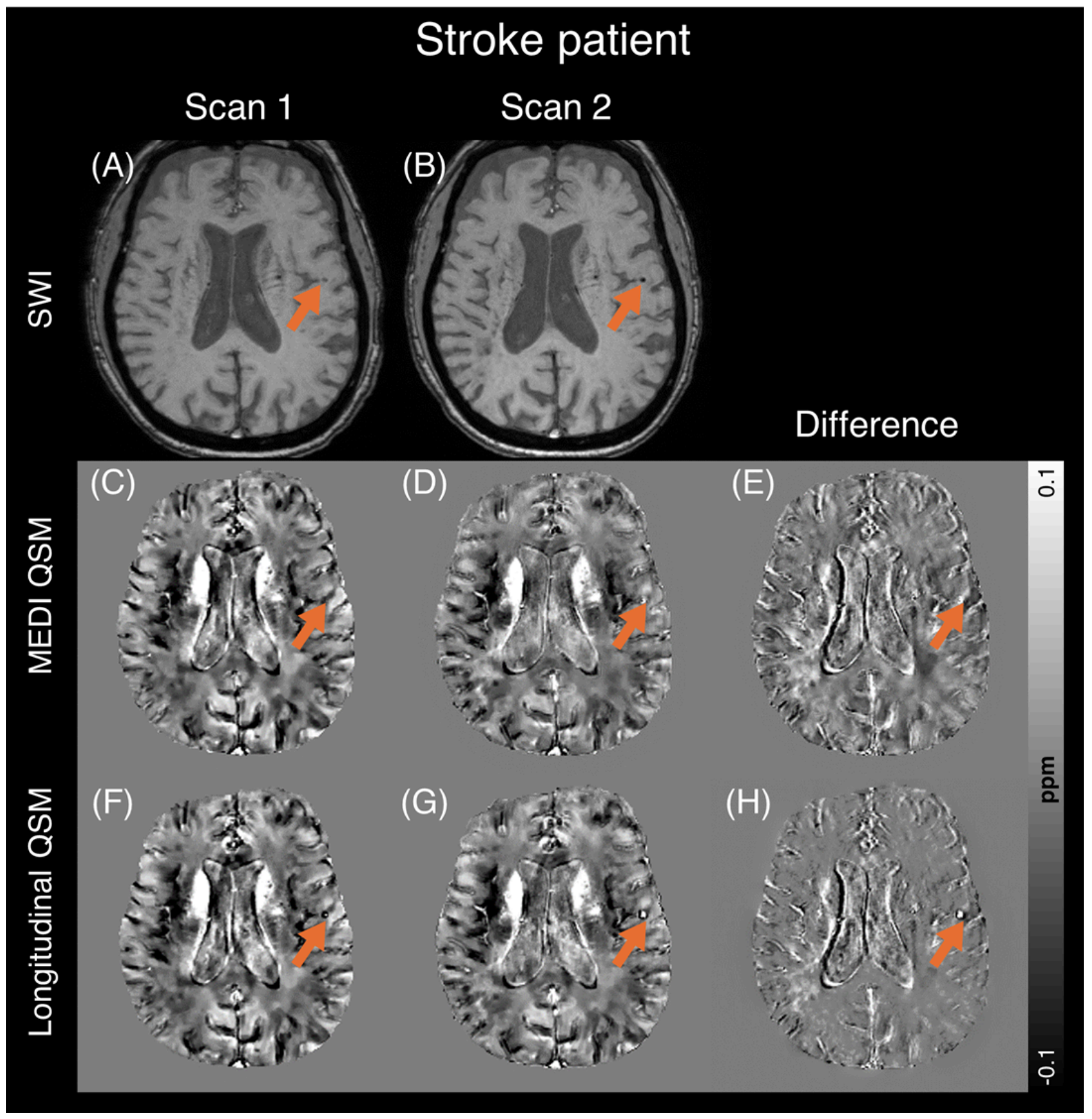

**Figure 7.** In-vivo results in the representative stroke patient. (A–B) SWIs from scan 1 and scan 2, acquired ten months apart, showing a lesion (orange arrows). (C–E) MEDI QSM reconstructions from the two scans, and their difference map. Widespread non-specific susceptibility variations are observed, which obscure lesion-related changes. (F–H) Longitudinal QSM reconstructions from the same scans and the corresponding difference map, showing clearer lesion depiction at scan 2 with more pronounced susceptibility changes compared with MEDI QSM.

## 3.4. In-vivo experiment: MS patients

Figure 8 shows longitudinal susceptibility maps from two representative MS patients reconstructed using MEDI QSM and Longitudinal QSM. Across both patients, MEDI QSM exhibits widespread susceptibility variations between scans that are not spatially confined to lesion regions (see the difference map). In contrast, Longitudinal QSM consistently suppresses non-lesion variability and yields more spatially localized inter-scan differences. As a result, temporal susceptibility changes are primarily confined to lesion regions in both patients, enabling clearer identification of lesion-specific evolution across scans.

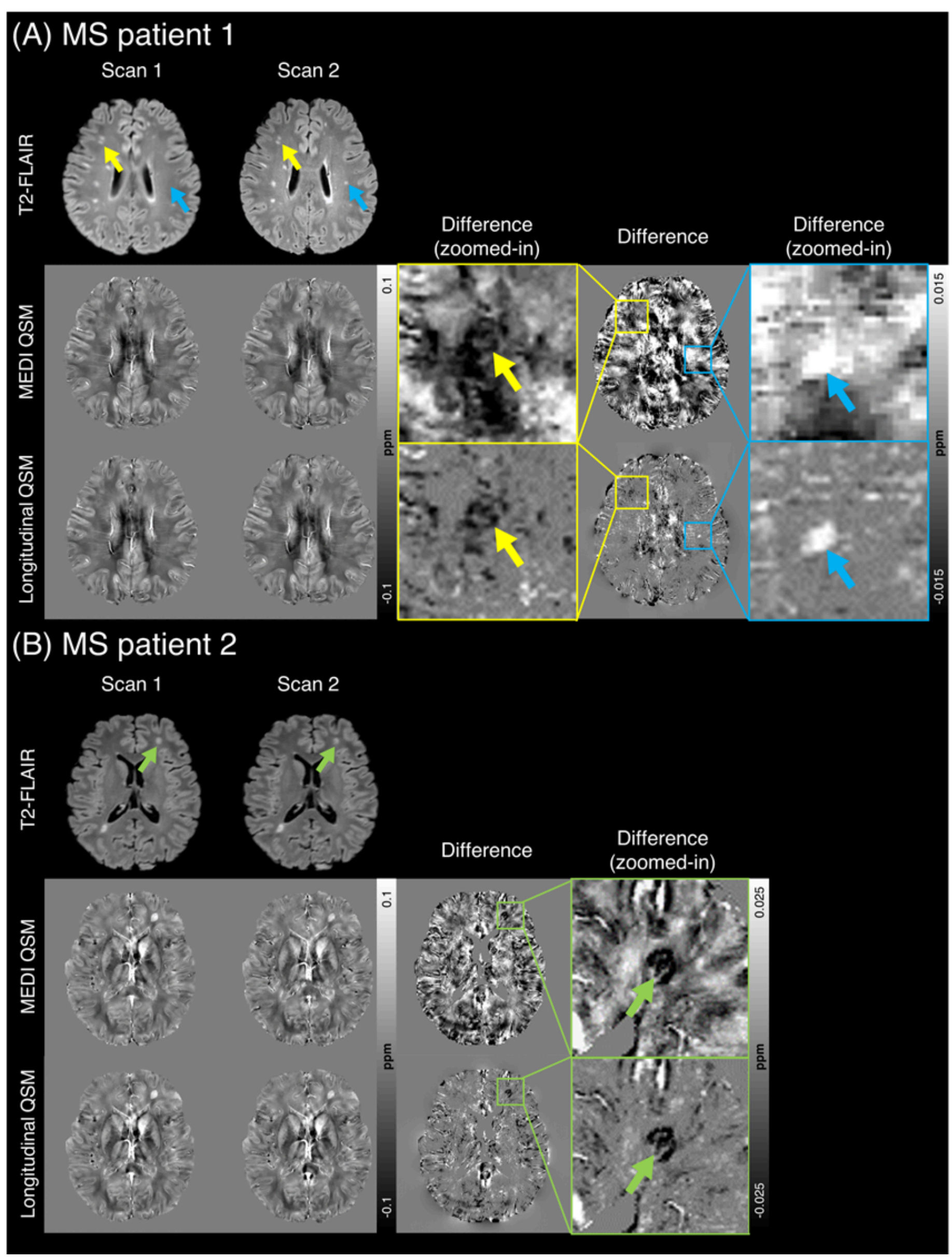

**Figure 8.** In-vivo results from two representative MS patients, (A) MS patient 1 and (B) MS patient 2. For each patient, scan 1 and scan 2 results are shown for T2-FLAIR, MEDI QSM, and the proposed Longitudinal QSM. Corresponding difference maps and zoomed-in views are provided for the QSM results. The inter-scan interval is eight months for patient 1 and fifteen months for patient 2. MEDI QSM exhibits widespread non-specific susceptibility variations that obscure lesion-related changes. In contrast, Longitudinal QSM suppresses background fluctuations and reveals clearer lesion-specific susceptibility changes (yellow and blue arrows in A, green arrows in B).

## 3.4. Ablation study

To assess the role of jointly fitting background susceptibility variations in Longitudinal QSM, an ablation study was performed. When temporal susceptibility differences are restricted to the brain region (brain-only setting), the resulting reconstructions exhibit increased temporal variability compared with the proposed framework, particularly near brain boundaries (Figure 9A). Across both scan–repeat and scan–rescan experiments in in-vivo healthy volunteers, the brain-only formulation results in higher NRMSE (35.6 ± 5.1% and 43.7 ± 9.0%, respectively) and lower SSIM (0.972 ± 0.01 and 0.953 ± 0.03, respectively) compared to the proposed method. ROI-wise variation also demonstrates higher values of ROI variations in the ablation setting (Figure 9B).

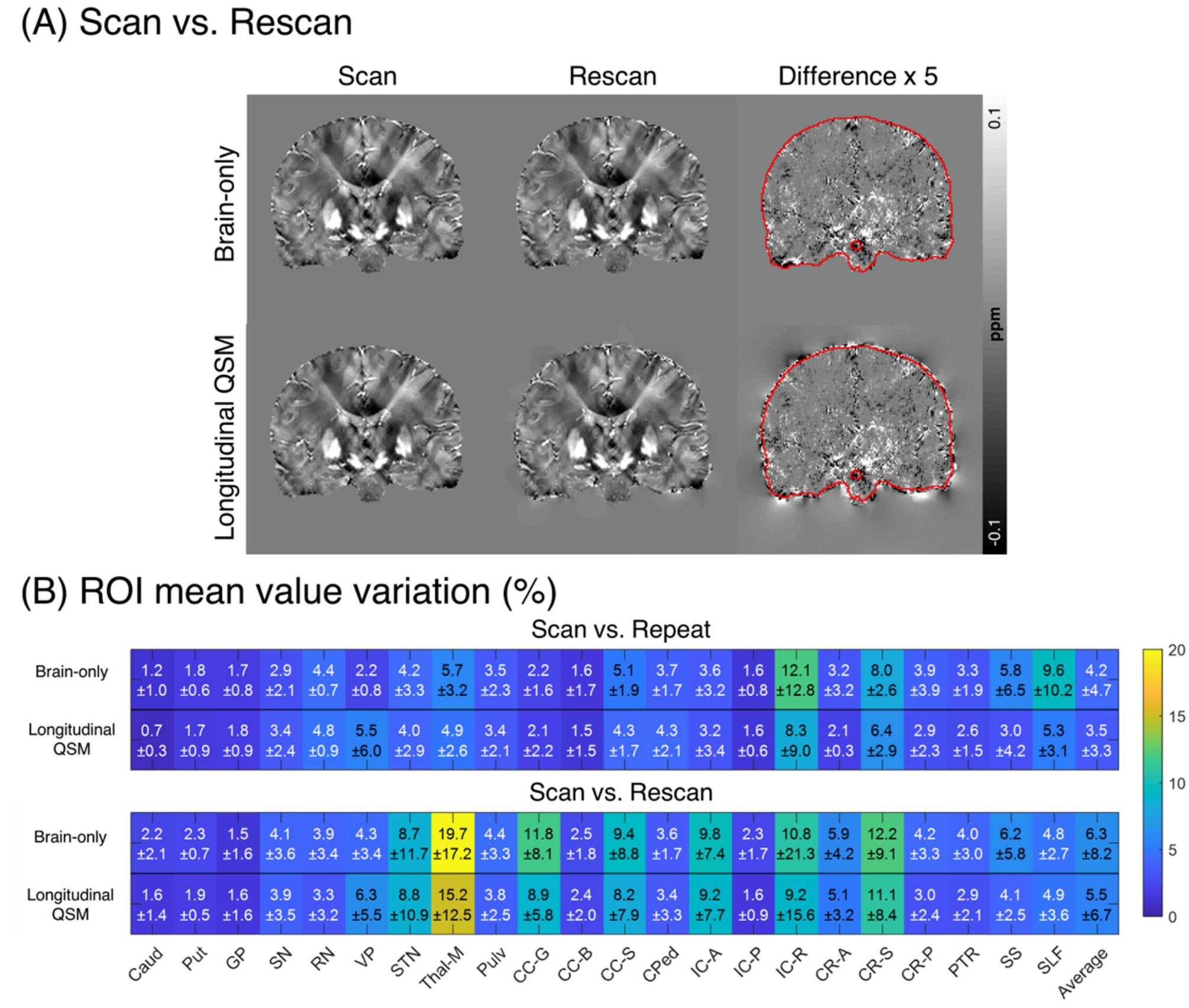

**Figure 9.** Ablation study results in in-vivo healthy dataset. (A) Representative in-vivo susceptibility maps for the scan–rescan experiment, comparing the brain-restricted ablation setting (brain-only) and the Longitudinal QSM. For each method, susceptibility maps from the two time points (scan and rescan) and their difference maps are shown. The red line indicates the brain mask. The ablation setting exhibits increased temporal variation near brain boundaries, whereas the proposed method shows reduced variation with more spatially consistent difference maps. (B) ROI-wise mean susceptibility variation (%) for scan–repeat and scan–rescan experiments. Values are reported as mean

± standard deviation across subjects. Overall, the proposed method demonstrates lower ROI-wise variability compared with the ablation setting. Caud: caudate nucleus, Put: putamen, GP: globus pallidus, NAc: nucleus accumbens, SN: substantia nigra, RN: red nucleus, VP: ventral pallidum, STN: subthalamic nucleus, Thal-M: medial thalamic nuclei, Thal-L: lateral thalamic nuclei, Pulv: pulvinar, CC-G: genu of corpus callosum, CC-B: body of corpus callosum, CC-S: splenium of corpus callosum, CPed: cerebral peduncle, IC-A: anterior limb of internal capsule, IC-P: posterior limb of internal capsule, IC-R: retrolenticular part of internal capsule, CR-A: anterior corona radiata, CR-S: superior corona radiata, CR-P: posterior corona radiata, PTR: posterior thalamic radiation, SS: sagittal stratum, SLF: superior longitudinal fasciculus.

Similar trends are observed in the simulation experiments, where the proposed method demonstrates improved comparable or improved voxel-wise and ROI-wise inter-scan consistency compared with the brain-only setting. In the scan–rescan simulation, the brain-only formulation resulted in NRMSE of 1.2 ± 0.2% and SSIM of 0.999 ± 0.001 in the no-rotation condition, 11.6 ± 1.5% and 0.997 ± 0.001 for the 3.4° rotation, and 14.7 ± 2.0% and 0.996 ± 0.001 for the 6.8° rotation. ROI mean susceptibility values also showed slightly larger variations in the brain-only setting, reaching an average of 0.8 ± 0.8% in the 6.8° rotation condition. These results indicate reduced inter-scan consistency relative to the proposed method.

For the lesion simulation, the brain-only formulation showed reduced sensitivity to true susceptibility changes, with a slope of 0.90 in the regression analysis. Correspondingly, the brain-only reconstruction resulted in relatively large NRMSE values, while the SSIM remained comparable to that of the proposed framework. Specifically, NRMSE values were 1.4 ± 0.2% in the no-rotation condition, 12.9 ± 1.5% for the 3.4° rotation, and 16.3 ± 2.1% for the 6.8° rotation, whereas the SSIM values were consistently high (0.999 ± 0.0001, 0.999 ± 0.0002, and 0.999 ± 0.0005, respectively). ROI-wise analysis further showed slightly larger susceptibility variations (0.8 ± 0.6%) in the ablation setting.

## 4. Discussion

This study introduces a simultaneous reconstruction framework for multiple time points QSM data that jointly estimates susceptibility maps across time points while accounting for background susceptibility variation and head orientation change. By incorporating a spatial sparsity constraint on temporal susceptibility differences within the brain, the method improves consistency between scans while maintaining sensitivity to localized susceptibility changes.

Through simulation (Figs. 2 and 3) and in-vivo experiments in healthy controls (Fig. 6), we demonstrate that even small changes in head orientation and background susceptibility can lead to substantial discrepancies between individually reconstructed QSM maps. These discrepancies are considerably suppressed by the proposed approach, indicating that simultaneous modeling of multiple time points can mitigate confounding factors that conventional independent reconstructions cannot correct.

In the simulated lesion experiments, we demonstrate that the proposed method not only improves longitudinal consistency but also preserves true susceptibility alterations (Figs. 4 and 5). In the conventional reconstruction, widespread susceptibility variations affect the entire brain, including the lesion region, making it difficult to distinguish true lesion-related changes from longitudinal measurement variability. In contrast, the proposed framework preserves the susceptibility change within the simulated lesion while suppressing unwanted variations. This suggests that the improved consistency achieved by the proposed method does not come at the cost of suppressing true susceptibility changes, but rather reflects a clearer separation from longitudinal measurement variability.

Application of the method to stroke (Figure 7) and MS patients (Figure 8) further illustrates its potential to capture lesion-related susceptibility changes. The approach reduces non-specific differences that appear in conventional reconstructions while preserving alterations in lesions, indicating improved sensitivity in detecting temporal changes. The proposed framework may therefore be beneficial for monitoring disease evolution, assessing treatment response, and improving the reliability of susceptibility-based biomarkers in clinical follow-up studies.

An ablation analysis clarifies the role of joint background susceptibility modeling in the proposed framework. When temporal susceptibility differences are constrained exclusively within the brain, reconstruction inconsistencies increase (Figure 9). This behavior indicates that background susceptibility variations propagate into the reconstructed brain susceptibility due to the imperfection of the dipole field deconvolution. Joint modeling of background susceptibility variations is therefore essential for preventing spurious temporal differences.

While our framework is evaluated on data with two time points, incorporating additional time points would enable characterization of more complex temporal trajectories and could further improve robustness by leveraging temporal redundancy. Such an extension would be particularly beneficial for studies requiring long-term monitoring.

The proposed framework assumes that temporal susceptibility alterations are spatially sparse. While this assumption is reasonable for many pathological processes, it can be violated in cases involving widespread susceptibility changes or pronounced anatomical changes, potentially leading to degraded performance.

In this study, sparsity is enforced on susceptibility differences between all time points. While this approach is tractable for a small number of time points, the number of sparsity terms increases approximately proportional to $N^2$ as the number of time points $N$ grows, potentially leading to higher computational complexity for $N > 4$. In such cases, alternative strategies, such as enforcing sparsity only on differences relative to the first time point or between adjacent time points, may be considered as they maintain a number of sparsity terms equal to $N-1$. These alternatives are equivalent to the all-to-all formulation when $N = 2$, and for $N = 3$ they involve one fewer sparsity term, with the computational advantage becoming more pronounced as $N$ increases. The impact of these alternative designs on reconstruction performance will be investigated in future studies using multiple time point data.

The proposed framework is developed on the MEDI QSM formulation [21], where the magnitude-guided edge prior helps preserve spatial structures at each time point. However, the key concept of this work is not limited to MEDI QSM itself, but rather lies in the simultaneous reconstruction of multiple time point susceptibility maps with sparsity enforced on their temporal differences. As such, the proposed framework can be extended to other QSM reconstruction methods. For example, a similar simultaneous reconstruction strategy could be implemented within iLSQR QSM reconstruction [28] by incorporating an additional sparsity regularizer on temporal differences. Furthermore, the framework could be extended to deep learning–based approaches [32], where a neural network may learn a dipole model. In such cases, the proposed temporal sparsity constraint could be integrated through an unrolled network architecture, enabling joint reconstruction while preserving sensitivity to localized temporal changes.

In the *in-vivo* healthy control results, the variability between time points is larger than that observed in simulations, while the rotation angles used in our simulations (3.4° and 6.8°) fall within the range of head orientation changes observed in vivo (4.8 ± 1.9°). This discrepancy may arise from factors not modeled in our simulation environment, such as noise, imperfect registration, and susceptibility anisotropy [33], [34]. Although the rotation angles in our study were small and thus anisotropy-related effects are expected to be modest, incorporating anisotropy into future simulations will help clarify its contribution. Further work is also necessary to quantify the influence of misregistration and noise.

Although the patient's case demonstrated that the proposed framework can reveal lesion-related susceptibility changes more clearly than conventional reconstructions, validation remains challenging because there is no ground-truth reference. Phantom experiments or synthetic datasets that mimic realistic longitudinal changes will be needed for further validation.

This work focuses on QSM, where ill-posed processing steps such as background field removal and dipole inversion amplify longitudinal variability. However, the proposed framework is generally applicable to other model-based reconstructions, such as parallel imaging reconstruction and diffusion MRI, where improving longitudinal consistency remains an important challenge.

## 5. Conclusion

In conclusion, this study presents a simultaneous reconstruction framework for multiple time point QSM data that jointly estimates susceptibility maps across time points while reducing variability caused by background susceptibility and head orientation changes. The method improves reproducibility and sensitivity to localized temporal changes in both simulations and in-vivo experiments. By providing more stable and robust longitudinal susceptibility measurements, the proposed framework offers a valuable tool for studies that require reliable tracking of subtle temporal susceptibility alterations, including investigations of neurodegenerative diseases as well as aging, development, and neuroplasticity.

## *References*

[1] A. Compston and A. Coles, “Multiple sclerosis,” The Lancet, vol. 372, no. 9648, pp. 1502–1517, 2008, doi: 10.1016/s0140-6736(08)61620-7.

[2] L. Zecca, M. B. H. Youdim, P. Riederer, J. R. Connor, and R. R. Crichton, “Iron, brain ageing and neurodegenerative disorders,” Nature Reviews Neuroscience, vol. 5, no. 11, pp. 863–873, 2004, doi: 10.1038/nrn1537.

[3] L. D. Rochefort, R. Brown, M. R. Prince, and Y. Wang, “Quantitative MR susceptibility mapping using piece-wise constant regularized inversion of the magnetic field,” Magnetic Resonance in Medicine, vol. 60, no. 4, pp. 1003–1009, 2008, doi: 10.1002/mrm.21710.

[4] K. Shmueli, J. A. D. Zwart, P. V. Gelderen, T. Q. Li, S. J. Dodd, and J. H. Duyn, “Magnetic susceptibility mapping of brain tissue in vivo using MRI phase data,” Magnetic Resonance in Medicine, vol. 62, no. 6, pp. 1510–1522, 2009, doi: 10.1002/mrm.22135.

[5] C. Langkammer et al., “Quantitative Susceptibility Mapping in Multiple Sclerosis,” Radiology, vol. 267, no. 2, pp. 551–559, 2013, doi: 10.1148/radiol.12120707.

[6] A. K. Lotfipour et al., “High resolution magnetic susceptibility mapping of the substantia nigra in Parkinson’s disease,” J. Magn. Reson. Imaging, vol. 35, no. 1, pp. 48–55, 2012, doi: 10.1002/jmri.22752.

[7] W. Kim et al., “χ-Separation Imaging for Diagnosis of Multiple Sclerosis versus Neuromyelitis Optica Spectrum Disorder.,” Radiology, vol. 307, no. 1, p. e220941, 2022, doi: 10.1148/radiol.220941.

[8] S. Ji et al., “Comparison between R2′-based and R2*-based χ-separation methods: A clinical evaluation in individuals with multiple sclerosis,” NMR Biomed., p. e5167, 2024, doi: 10.1002/nbm.5167.

[9] V. Zachariou, C. Pappas, C. E. Bauer, E. R. Seago, and B. T. Gold, “Exploring the links among brain iron accumulation, cognitive performance, and dietary intake in older adults: A longitudinal MRI study,” Neurobiol. Aging, vol. 145, pp. 1–12, 2025, doi: 10.1016/j.neurobiolaging.2024.10.006.

[10] G. E. C. Thomas, N. Hannaway, A. Zarkali, K. Shmueli, and R. S. Weil, “Longitudinal Associations of Magnetic Susceptibility with Clinical Severity in Parkinson’s Disease,” Mov. Disord., vol. 39, no. 3, pp. 546–559, 2024, doi: 10.1002/mds.29702.

[11] Y. Zhang et al., “Longitudinal change in magnetic susceptibility of new enhanced multiple sclerosis (MS) lesions measured on serial quantitative susceptibility mapping (QSM),” J. Magn. Reson. Imaging, vol. 44, no. 2, pp. 426–432, 2016, doi: 10.1002/jmri.25144.

[12] J. Hagemeier et al., “Evolution of Brain Iron Levels in Multiple Sclerosis: A 2-Year Longitudinal Quantitative Susceptibility Mapping Study at 3T (P4.163),” Neurology, vol. 86, no. 16_supplement, 2016, doi: 10.1212/wnl.86.16_supplement.p4.163.

[13] H.-G. Shin et al., “Association of iron deposition in MS lesion with remyelination capacity using susceptibility source separation MRI,” NeuroImage: Clin., vol. 45, p. 103748, 2025, doi: 10.1016/j.nicl.2025.103748.

[14] J. Müller et al., “Quantifying Remyelination Using χ-Separation in White Matter and Cortical Multiple Sclerosis Lesions,” Neurology, vol. 103, no. 6, p. e209604, 2024, doi: 10.1212/wnl.0000000000209604.

[15] K. Deh et al., “Reproducibility of quantitative susceptibility mapping in the brain at two field strengths

from two vendors," J. Magn. Reson. Imaging, vol. 42, no. 6, pp. 1592–1600, 2015, doi: 10.1002/jmri.24943.
[16] M. Lancione et al., "Multi-centre and multi-vendor reproducibility of a standardized protocol for quantitative susceptibility Mapping of the human brain at 3T," Physica Medica, vol. 103, pp. 37–45, 2022, doi: 10.1016/j.ejmp.2022.09.012.
[17] C. Rua et al., "Multi-centre, multi-vendor reproducibility of 7T QSM and R2* in the human brain: Results from the UK7T study," Neuroimage, vol. 223, p. 117358, 2020, doi: 10.1016/j.neuroimage.2020.117358.
[18] O. C. Kiersnowski, A. Karsa, S. J. Wastling, J. S. Thornton, and K. Shmueli, "Investigating the effect of oblique image acquisition on the accuracy of QSM and a robust tilt correction method," Magn. Reson. Med., vol. 89, no. 5, pp. 1791–1808, 2023, doi: 10.1002/mrm.29550.
[19] Z. Liu, P. Spincemaille, Y. Yao, Y. Zhang, and Y. Wang, "MEDI+0: Morphology enabled dipole inversion with automatic uniform cerebrospinal fluid zero reference for quantitative susceptibility mapping," Magnetic Resonance in Medicine, vol. 79, no. 5, pp. 2795–2803, 2018, doi: 10.1002/mrm.26946.
[20] T. Kim et al., "Vessel segmentation for χ$$ \chi $$-separation in quantitative susceptibility mapping," Magn. Reson. Med., 2025, doi: 10.1002/mrm.70054.
[21] T. Liu et al., "Morphology enabled dipole inversion (MEDI) from a single-angle acquisition: Comparison with COSMOS in human brain imaging," Magn. Reson. Med., vol. 66, no. 3, pp. 777–783, 2011, doi: 10.1002/mrm.22816.
[22] I. Daubechies, R. DeVore, M. Fornasier, and C. S. Güntürk, "Iteratively reweighted least squares minimization for sparse recovery," Commun. Pure Appl. Math., vol. 63, no. 1, pp. 1–38, 2010, doi: 10.1002/cpa.20303.
[23] T. Liu, P. Spincemaille, L. de Rochefort, B. Kressler, and Y. Wang, "Calculation of susceptibility through multiple orientation sampling (COSMOS): A method for conditioning the inverse problem from measured magnetic field map to susceptibility source image in MRI," Magn. Reson. Med., vol. 61, no. 1, pp. 196–204, 2009, doi: 10.1002/mrm.21828.
[24] M. Kim et al., "χ-sepnet: Deep Neural Network for Magnetic Susceptibility Source Separation," Hum. Brain Mapp., vol. 46, no. 2, p. e70136, 2025, doi: 10.1002/hbm.70136.
[25] M. Jenkinson, C. F. Beckmann, T. E. J. Behrens, M. W. Woolrich, and S. M. Smith, "FSL," NeuroImage, vol. 62, no. 2, pp. 782–790, 2012, doi: 10.1016/j.neuroimage.2011.09.015.
[26] B. Dymerska et al., "Phase unwrapping with a rapid opensource minimum spanning tree algorithm (ROMEO)," Magn. Reson. Med., vol. 85, no. 4, pp. 2294–2308, 2021, doi: 10.1002/mrm.28563.
[27] Wu, W. Li, A. V. Avram, S.-M. Gho, and C. Liu, "Fast and tissue-optimized mapping of magnetic susceptibility and T2* with multi-echo and multi-shot spirals," NeuroImage, vol. 59, no. 1, pp. 297–305, 2012, doi: 10.1016/j.neuroimage.2011.07.019.
[28] W. Li et al., "A method for estimating and removing streaking artifacts in quantitative susceptibility mapping," NeuroImage, vol. 108, pp. 111–122, 2015, doi: 10.1016/j.neuroimage.2014.12.043.
[29] K. Min et al., "A human brain atlas of χ-separation for normative iron and myelin distributions," NMR Biomed., p. e5226, 2024, doi: 10.1002/nbm.5226.
[30] F. Schweser, A. Deistung, B. W. Lehr, and J. R. Reichenbach, "Quantitative imaging of intrinsic magnetic tissue properties using MRI signal phase: An approach to in vivo brain iron metabolism?," NeuroImage, vol. 54,

no. 4, pp. 2789–2807, 2011, doi: 10.1016/j.neuroimage.2010.10.070.
[31] Wu, W. Li, A. Guidon, and C. Liu, “Whole brain susceptibility mapping using compressed sensing,” Magnetic Resonance in Medicine, vol. 67, no. 1, pp. 137–147, 2012, doi: 10.1002/mrm.23000.
[32] J. Yoon et al., “Quantitative susceptibility mapping using deep neural network: QSMnet,” NeuroImage, vol. 179, pp. 199–206, 2018, doi: 10.1016/j.neuroimage.2018.06.030.
[33] J. Lee et al., “Sensitivity of MRI resonance frequency to the orientation of brain tissue microstructure,” Proceedings of the National Academy of Sciences of the United States of America, vol. 107, no. 11, pp. 5130–5135, 2010, doi: 10.1073/pnas.0910222107.
[34] J. Lee, H.-G. Shin, W. Jung, Y. Nam, S.-H. Oh, and J. Lee, “An R2* model of white matter for fiber orientation and myelin concentration,” NeuroImage, vol. 162, pp. 269–275, 2017, doi: 10.1016/j.neuroimage.2017.08.050.